\documentclass[conference]{IEEEtran}

\IEEEoverridecommandlockouts                              

\usepackage{graphicx}
\usepackage{epstopdf}
\usepackage{graphics} 
\usepackage{caption}
\usepackage{epsfig} 
\usepackage{color}
\usepackage{amsmath}
\usepackage{amsthm}
\usepackage{amssymb} 
\usepackage{verbatim}  
\usepackage{mathtools}
\usepackage{subfig}
\usepackage{tabularx}
\usepackage{verbatim}  
\newtheorem{theorem}{Theorem}
\newtheorem{lemma}{Lemma}
\newtheorem{proposition}{Proposition}

\newtheorem{remark}{Remark}
\usepackage[ colorlinks = true,
linkcolor = blue,
urlcolor = blue,
citecolor = red,
anchorcolor = green,
]{hyperref}

\newcommand{\bp}{\mathbf{p}}
\newcommand{\bW}{\mathbf{W}}
\newcommand{\bw}{\mathbf{w}}

\newcommand{\bPsi}{\boldsymbol{\Psi}}

\allowdisplaybreaks[1]

\title{Hypothesis Testing under Maximal Leakage Privacy Constraints}
\author{\IEEEauthorblockN{Jiachun Liao, Lalitha Sankar}
\IEEEauthorblockA{School of Electrical, Computer\\ and Energy Engineering,\\
		Arizona State University\\
		Email: \{jiachun.liao,lalithasankar\}@asu.edu}
\and
\IEEEauthorblockN{Flavio P. Calmon}
\IEEEauthorblockA{IBM T.J.~Watson Research Center\\
	 in Yorktown Heights, New York\\
Email: fdcalmon@us.ibm.com\\
}
\and
\IEEEauthorblockN{Vincent Y. F. Tan}
\IEEEauthorblockA{Department of ECE, \\
	Department of Mathematics,\\
	 National University of Singapore \\
	  Email: {vtan@nus.edu.sg}
 }
\thanks{This work is supported in part by the National Science Foundation under grants CCF\--1350914 and CIF\--1422358.}
}

\begin{document}
	\maketitle	
	
\begin{abstract}
		The problem of publishing privacy-guaranteed data for hypothesis testing is studied using the maximal leakage (ML) as a metric for privacy and the type-II error exponent as the utility metric. The optimal mechanism (random mapping) that maximizes utility for a bounded leakage guarantee is determined for the entire leakage range for binary datasets. For non-binary datasets, approximations in the high privacy and high utility regimes are developed. The results show that, for any desired leakage level, maximizing utility forces the ML privacy mechanism to reveal partial to complete knowledge about a subset of the source alphabet. The results developed on maximizing a convex function over a polytope may also of an independent interest.
\end{abstract}

\section{Introduction}
There is tremendous value to publishing datasets for a variety of statistical inference applications; however, it is crucial to ensure that the published dataset while providing utility does not reveal other information. Specifically, the published dataset should allow the intended inference to be made while limiting other inferences. This requires using a randomizing mechanism (i.e., a \textit{noisy channel}) that guarantees a certain measure of privacy; however, any such privacy mechanism will, in turn, reduce the fidelity of the intended inference leading to a tradeoff between utility of the published data and the privacy of the respondents in the dataset. 

Recently, in \cite{MaximalLeakage_Issa2016}, Issa \textit{et al.} (see also \cite{ML_Alvim2012}) propose the metric of maximal leakage (ML) as a measure of the gain, in bits, in guessing any function of the original data from the published data and show that it is effectively the Sibson mutual information of order $\infty$ of the randomized mapping (the privacy mechanism) from the alphabet of the original to that of the published dataset. Inspired by the operational significance of the ML metric, for the statistical application of binary hypothesis testing, we determine the privacy-utility tradeoff (PUT) using ML as the privacy metric and the type-II error exponent as the utility metric. We consider the \textit{local privacy} model in which the same (memoryless) mechanism is applied independently to each entry of the dataset. This captures a large class of applications in which the respondents of a dataset can apply a privacy mechanism before sharing data. We present closed-form expressions for the PUT for binary sources for any leakage level; for arbitrary source alphabets we obtain the PUT in the high and low privacy regimes. 

Our results show that the maximal leakage mechanism in trying to simultaneously  ensure utility and restrict leakage yields output symbols that reveal either complete or partial knowledge about a subset of the source symbols. Specifically for the binary case, the mechanism is such that the published data reveals the original data for one of the source letters without any uncertainty! This behavior results from a combination of the maximal leakage requirement (feasible polytope) and the convexity of the relative entropy as a function of the privacy mechanism (randomized mapping). 

  \textbf{Notation:}
We use bold capital letters to represent matrices, e.g. $\mathbf{X}$ is a matrix whose $i^{\mathrm{th}}$ column is $\mathbf{X}_i$ and $(i,j)^{\mathrm{th}}$ entry is $X_{ij}$. We use bold lower case letters to represent \textbf{row} vectors, e.g. $\mathbf{x}$ is a vector with the $i^{\mathrm{th}}$ entry $x_i$. We denote sets by capital calligraphic letters, e.g., $\mathcal{X}$.
For a vector $\mathbf{x}$ with entries $x_i$, $[\mathbf{x}]$ is a {\em diagonal} matrix whose the $i^{\mathbf{th}}$ diagonal entry is $x_i$.   We use $\|\cdot \|$ and $\log(\cdot) $ to denote the $\ell_2$ norm and logarithm with base $2$, respectively; $D$ denotes the relative entropy. The phase ``column permutation'' implies the application of a permutation operation on the columns of a matrix.

\section{Problem Formulation} 
Binary hypothesis testing is a statistical inference problem concerning the decision  between two distinct probability distributions of the observed data. Let $X^n=(X_1,X_2,\ldots,X_n)$ denote a sequence of $n$ random variables, whose entries $X_i\in \mathcal{X}, i \in \{1,2,\ldots,n\}$, are independent and identically distributed (i.i.d.) according to a distribution $\mathbf{p}$ that is hypothesized to be either $H_1:\, \mathbf{p}=\mathbf{p}_1$ or $H_2:\, \mathbf{p}=\mathbf{p}_2$. Let $\beta^{(n)}_1$ and $\beta^{(n)}_2$ be the probabilities of error, such that type-I error $\beta^{(n)}_1$ (resp.\ type-II error $\beta^{(n)}_2$) is the probability of choosing $H_2$ (resp. $H_1$) when original data $\mathbf{x}^n$ is generated by $\mathbf{p}_1$ (resp.\ $\mathbf{p}_2$). From the Chernoff-Stein lemma~\cite[Chap.~11]{IT_Cover}, under a constraint that  $\beta^{(n)}_1 \in (0,1)$, the maximal asymptotic error exponent of $\beta^{(n)}_2$ is $D(\mathbf{p}_1 \| \mathbf{p}_2)$. The relative entropy $D(\mathbf{p}_1 \| \mathbf{p}_2)$ is a measure of the accuracy of hypothesis testing.

For i.i.d.\ datasets considered here, we restrict our analysis to memoryless \textit{privacy mechanisms}. A memoryless privacy mechanism independently maps entry $X_i\sim\bp$ of $X^n$ to an output $\hat{X}_i\in\hat{\mathcal{X}}$ to obtain a released dataset $\hat{X}^n$. Formally, a privacy mechanism $\bW$ is an $M\times M$ row-stochastic conditional probability matrix with entries $W_{ij}=\Pr\{\hat{X}=j|X=i\}$, $i,j\in\{1,\ldots,M\}$. As a result of applying a privacy mechanism, the hypothesis test is now performed on the i.i.d.\ sequence $\hat{X}^n$ distributed as either $\mathbf{p}_1\mathbf{W}$ or $\mathbf{p}_2\mathbf{W}$. It is easy to see that the resulting   type-II error exponent of the test is $D(\mathbf{p}_1\mathbf{W} \| \mathbf{p}_2\mathbf{W})$. We choose this function (of $\mathbf{W}$) as the utility of the privacy-guaranteed hypothesis test.

For random variables $X$ and $\hat{X}$ as well as the privacy mechanism $\bW$ defined above, the following proposition summarizes the definition and simplification of  the maximal leakage in {\cite[Def.~1, Thm.~1, Cor.~1]{MaximalLeakage_Issa2016}}.
\begin{proposition}
	Given a joint distribution $[\bp]\bW$ of $(X,\hat{X})$, the maximal leakage from $X$ to $\hat{X}$ is defined as
	\begin{align}\label{eq:PracticalDefine_ML}
	L(X\rightarrow \hat{X})=\sup_{\substack{U\--X\--\hat{X}\--\hat{U}}} \log \frac{\Pr\{U=\hat{U}\}}{\max_{u\in\mathcal{U}}\Pr\{U=u\}}.
	\end{align}
	The expression in \eqref{eq:PracticalDefine_ML} is equivalent to 
	\begin{align}\label{eq:Define_ML}
	L(X\rightarrow \hat{X})=\log \Big(\sum_{j=1}^{M} \max_i\{W_{ij}\}\Big)=I_{\infty}(X;\hat{X}).
	\end{align}
\end{proposition}
\begin{remark}
	In the sequel, we assume that $\bp_1$ and $\bp_2$ have the same (and full) support. Under this assumption, we note that the expression in \eqref{eq:Define_ML} is independent of the source distribution. In particular, the single constraint on the $M^2$ entries of $\bW$ in \eqref{eq:Define_ML} suggests that multiple mechanisms can achieve the same leakage. However, these mechanisms will generally have different implications with respect to privacy protection. For example, for $M=4$, the following two matrices, $\bW_1$ and $\bW_2$, have the same maximal leakage of $1\text{ bit}$. 
	\begin{align}
	\bW_1=\begin{bmatrix}
	1 & 0 & 0 & 0\\0 & 0.3 & 0.3 & 0.4\\0 & 0.3 & 0.3 & 0.4\\0 & 0.3 & 0.3 & 0.4
	\end{bmatrix}\quad \bW_2=\begin{bmatrix}
	0.5 & 0.5 & 0 & 0\\0.5 & 0.5 & 0 & 0\\ 0 & 0 & 0.5 & 0.5\\0 & 0 & 0.5 & 0.5
	\end{bmatrix}\nonumber
	\end{align}
	However, for $\bW_2$, given an output symbol, there is some uncertainty regarding the input symbol. In contrast, for $\bW_1$, one of the output symbol completely reveals the corresponding input symbol.
\end{remark}

\begin{lemma}\label{Lemma:Maximal_Leakage}
	The function $I_{\infty}(X;\hat{X})\triangleq I_{\infty}(\bp,\bW)$ satisfies the following properties:
	\begin{itemize}
		\item[1.] $0\leq I_{\infty}(\bp,\bW)\leq \log M$;
		\item[2.] $I_{\infty}(\bp,\bW)=0$  $\Leftrightarrow$ $\sum_{j}\max_i\{W_{ij}\}=1$ $\Leftrightarrow$ $\bW$ is rank-1;
		\item[3.] $I_{\infty}(\bp,\bW)=\log M$ $\Leftrightarrow$ $\mathbf{W}$ is a permutation of identity matrix $\mathbf{I}$.
	\end{itemize}
\end{lemma}
The proof of Lemma \ref{Lemma:Maximal_Leakage} is in Appendix \ref{proof:Lemma:Maximal_Leakage}. Basically these properties follow from \eqref{eq:Define_ML}.

\subsection{Privacy\--Utility Trade-off}
The PUT for the binary hypothesis testing problem with maximal leakage and relative entropy as privacy and utility measures, respectively, is given by the following non-convex optimization 
\begin{subequations}\label{eq:MaximalLeakage_original}
	\begin{align}
	\max_{\substack{\bW}} \quad & D(\bp_1\bW\|\bp_2\bW)\nonumber\\
	\label{eq:MaximalLeakage_Constraint}
	\text{s.t.} \quad & \sum_{j=1}^{M} \max_{i} W_{ij}\leq 2^l\\
	\label{eq:MaximalLeakage_ConstraintW1}
	&\sum_{j=1}^{M}W_{ij}=1 \qquad\, \text{for all }i\\
	\label{eq:MaximalLeakage_ConstraintW2}
	& W_{ij}\geq 0 \qquad \qquad\text{for all }i,j
	\end{align}
\end{subequations}
where $l\in [0, \log M]$. 

By adding $M$ slack variables $\epsilon_j$, $j\in\{1,\ldots,M\}$, the privacy constraint in \eqref{eq:MaximalLeakage_Constraint} can be rewritten as
\begin{align}
\label{eq:MaximalLeakage_Constraint_rewritten1}
W_{ij}&\leq \epsilon_j \quad \text{ for all }i,j \in\{1,\ldots,M\}\\
\label{eq:MaximalLeakage_Constraint_rewritten2}
\sum_{j=1}^{M}\epsilon_j &\leq 2^l.
\end{align}
From \eqref{eq:MaximalLeakage_Constraint_rewritten1} and \eqref{eq:MaximalLeakage_Constraint_rewritten2} in conjunction with \eqref{eq:MaximalLeakage_ConstraintW1} and \eqref{eq:MaximalLeakage_ConstraintW2}, we note that the feasible region of \eqref{eq:MaximalLeakage_original} is a $M^2+M$~dimensional polytope resulting from $2M^2+M+1$ linear constraints. 
The optimization problem is non-convex since it involves maximizing a convex function. Furthermore, since the feasible region is a polytope, the optimal solutions are on the boundary. Specifically, at least one corner point of the polytope is an optimal solution of \eqref{eq:MaximalLeakage_original}. However, enumerating the vertices of the polytope is infeasible. As a first step to obtaining closed-form solutions, in the following theorem, we highlight properties of the optimal solutions.
\begin{theorem}\label{theorem:ML_OptSolChara}
	For an optimal solution $\bW^*$ of \eqref{eq:MaximalLeakage_original}, all column permutations of $\bW^*$ are also optimal solutions. If $\bW^*$ has at least one zero column, an infinite number of solutions are optimal.
\end{theorem}
A detailed proof is in Appendix \ref{proof:theorem:ML_OptSolChara}, 
and we briefly highlight the intuition here. Optimality of all permutations of $\bW^*$ follows from the fact that the column permutation of $\bW^*$ preserves both the objective and the constraints. Furthermore, when $\bW^*$ has at least one all-zero column, convex combinations of $\bW^*$ and any column permutation of $\bW^*$ that involves permuting one of the all-zero columns are also optimal. For example, let
\begin{align}\nonumber
	 \bW_1=\begin{bmatrix}
	 0 &1-a & a \\0  &1-b &b\\ 0  &1-c &c
	 \end{bmatrix}\quad \text{and}\quad \bW_2=\begin{bmatrix}
	 1-a & 0 & a \\1-b & 0 & b \\1-c & 0 & c
	 \end{bmatrix}.
\end{align}
For the maximal leakage $l=1+\max\{a,b,c\}-\min\{a,b,c\}$, where $a,b,c \in (0,1)$, if $\bW_1$ is an optimal solution of \eqref{eq:MaximalLeakage_original}, then so is $\bW_2$, and so are all convex combinations of $\bW_1$ and $\bW_2$, since they preserve the objective.

In the following section, we obtain a closed-form expression for the PUT in \eqref{eq:MaximalLeakage_original} for binary sources.

\section{Trade-off for Binary Sources}
Without loss of generality, for binary sources, the probability distributions $\bp_1$ and $\bp_2$ can be represented by Bernoulli parameters $p_1$ and $p_2$. Furthermore, note that $l\in [0,1]$. Define 
\begin{align}
	&f_1(p_1,p_2,l)\nonumber\\
	&\triangleq(p_{1}-1)(2^l-1)\log \frac{(1-p_{2})\big((2-2^l)+p_{1}(2^l-1)\big)}{(1-p_{1})\big((2-2^l)+p_{2}(2^l-1)\big)}\nonumber\\
					\label{eq:binary_optimalvalue_point1}
	&\quad+\log \frac{(2-2^l)+p_{1}(2^l-1)}{(2-2^l)+p_{2}(2^l-1)}\\
		\label{eq:binary_optimalvalue_point2}
	&f_2(p_1,p_2,l)\nonumber\\
	&\triangleq p_{1}(2^l-1)\log \frac{p_{1}\big(1+p_{2}(1-2^l)\big)}{p_{2}\big(1+p_{1}(1-2^l)\big)}+\log \frac{1+p_{1}(1-2^l)}{1+p_{2}(1-2^l)}.
\end{align}
\begin{theorem}\label{theorem:ML_binarysource}
	For binary hypotheses $H_1: p=p_1$ and $H_2: p=p_2$, and a chosen maximal leakage $l \in[0,1]$, the maximal utility (error exponent) is given by 
	\begin{align}
		\max\big\{f_1(p_1,p_2,l),f_2(p_1,p_2,l)\big\}
	\end{align}
	and is achieved by 
	\begin{align}\label{eq:Binary_optimalW}
        \bW^*=\begin{cases}
		\begin{bmatrix}
		2-2^l & 2^l-1\\
		1 & 0
		\end{bmatrix} 
		\text{ or }	
		\begin{bmatrix}
		2^l-1 & 2-2^l\\
		0 & 1
		\end{bmatrix}, \\
		\text{ if } f_1(p_1,p_2,l)\geq f_2(p_1,p_2,l)\\
		\\
		\begin{bmatrix}
				0 & 1\\
				2^l-1 & 2-2^l
		\end{bmatrix}
				\text{ or }	
		\begin{bmatrix}
			1 & 0\\
			2-2^l & 2^l-1	
		\end{bmatrix} . \\
		\text{ if } f_1(p_1,p_2,l)\leq f_2(p_1,p_2,l)\\
		\end{cases}
	\end{align}
\end{theorem}
\begin{figure}[t]
		\centering\includegraphics[width=3.3in]{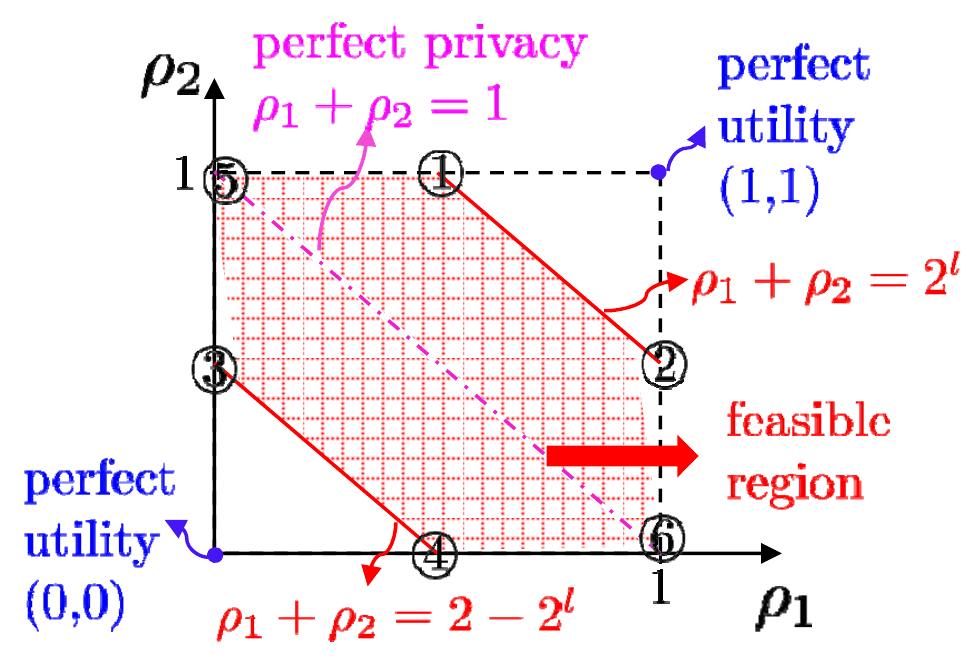}
		\caption{Feasible region of for binary sources: $\rho_1$ and $\rho_2$ are the off-diagonal entries of privacy mechanism $\bW$}
		\label{fig:ML_binaryregion}
		\vspace{-0.5cm}
\end{figure}
Due to space restrictions, we briefly outline the proof 
(see Appendix \ref{proof:theorem:ML_binarysource} for details).
Let $\rho_1$ and $\rho_2$ be the off-diagonal entries of the privacy mechanism $\bW$. The constraint in \eqref{eq:MaximalLeakage_Constraint} simplifies to 
	\begin{align}
	2-2^l\leq \rho_1+\rho_2\leq 2^l
	\end{align}
	and along with these in \eqref{eq:MaximalLeakage_ConstraintW1} and \eqref{eq:MaximalLeakage_ConstraintW2} yields the shaded region in the $(\rho_1,\rho_2)$ space as shown in Fig.~\ref{fig:ML_binaryregion}. The vertices \raisebox{.5pt}{\textcircled{\raisebox{-.9pt} {$1$}} and \textcircled{\raisebox{-.9pt} {$4$}}} result from column permutations of $\bW$, and thus, have the same utility. Similarly, vertices \raisebox{.5pt}{\textcircled{\raisebox{-.9pt} {$2$}} and \textcircled{\raisebox{-.9pt} {$3$}}} have the same utility. On the other hand, vertices \raisebox{.5pt}{\textcircled{\raisebox{-.9pt} {$5$}} and \textcircled{\raisebox{-.9pt} {$6$}}} achieve zero utility. Therefore, it suffices to compare the utilities at vertices \raisebox{.5pt}{\textcircled{\raisebox{-.9pt} {$1$}} and \textcircled{\raisebox{-.9pt} {$2$}}} given by $f_1$ in \eqref{eq:binary_optimalvalue_point1} and $f_2$ in \eqref{eq:binary_optimalvalue_point2}, respectively.
	
    From \eqref{eq:Binary_optimalW}, we note that for all $l$ and both cases ($f_1\geq f_2$ or $f_1\leq f_2$), there is no uncertainty of the input symbol given one of the output symbols. This is a direct consequence of the convexity of the relative entropy and the linearity of the ML constraint. These observations, coupled with fact that $2^l\geq 1$ for the binary case, forces one of the $\rho_i$ to be $1$ (or $0$).

\section{Trade-off for Arbitrary Alphabets}
In this section, we consider non-binary sources, i.e., $M>2$. Because it is challenging to find closed-form solutions, we focus on two extremal regimes: namely, the high privacy ($l\approx 0$) low utility and low privacy ($l\approx\log M$) high utility regimes. In each regime, we exploit the continuous differentiability  of $D$ in $\mathbf{W}$ to approximate the relative entropy objective about the extremal points (presented in Lemma \ref{Lemma:Maximal_Leakage}). This allows us to simplify the optimization problem and subsequently obtain closed-form solutions.
\subsection{Euclidean Approximation in High Privacy regime}
From Lemma \ref{Lemma:Maximal_Leakage}, recall that  for the perfect privacy case, i.e., $l=0$, the optimal mechanism $\bW_0$ is a rank-1 row stochastic matrix. In particular, all rows of $\bW_0$ are the same vector $\bw_0$. Thus the two output distributions are the same, i.e., $\bp_k\bW_0=\bw_0$ for all $k=1,2$. In the high privacy regime, we introduce an Euclidean information theoretic (EIT) approximation (see also  \cite{EITzheng2008}, \cite{EIT2015}) of $D(\bp_1\bW\|\bp_2\bW)$ by restricting its Taylor series about $\mathbf{W}=\bW_0$ to the second (quadratic) term $\frac{1}{2}\big\|(\mathbf{p}_1-\mathbf{p}_2)\bW[(\mathbf{w}_0)^{-\frac{1}{2}}]\big\|^2$ (see also \cite{Liao_Allerton16}). The mechanism $\bW$ is assumed to be in a small neighborhood about $\bW_0$, 
such that for some small $\delta\in[0,\frac{1}{M}]$, $|(\bp\bW)_j-w_{0j}|\leq \delta $ for all $j$. This, in turn, implies that $l\leq \log(1+M\delta)$.
Therefore, for $l\in [0, 1]$, the EIT approximation of the utility function $D(\bp_1\bW\|\bp_2\bW)$ results in the following optimization%
\begin{equation}\label{eq:ML_HPapproximation}
	\begin{aligned}
		\max_{\substack{\bW}} \quad & \frac{1}{2}\big\|(\mathbf{p}_1-\mathbf{p}_2)\bW[(\mathbf{w}_0)^{-\frac{1}{2}}]\big\|^2\\
		\text{s.t.} \quad 
		& \sum_{j=1}^{M} \max_{i} W_{ij}\leq 2^l\\
		&\sum_{j=1}^{M}W_{ij}=1 \qquad\, \text{for all }i\\
		&  W_{ij}\geq 0 \qquad \qquad\text{for all }i,j.\\
	\end{aligned}
\end{equation}
Since the feasible regions in \eqref{eq:ML_HPapproximation} and \eqref{eq:MaximalLeakage_original} are the same, the optimal solutions of the EIT approximation in \eqref{eq:ML_HPapproximation} are also feasible for the original PUT in \eqref{eq:MaximalLeakage_original}, thus the utility of the optimal solutions of \eqref{eq:ML_HPapproximation} is a lower bound of the optimal utility of \eqref{eq:MaximalLeakage_original}, and the EIT approximation is tight for $l$ very close to $0$, i.e., $\delta\approx 0$.

Observe that in \eqref{eq:ML_HPapproximation}, the objective depends on $\bp_1-\bp_2$ and being a convex function, is maximized on the boundary of the feasible region. Let $\mathcal{I}_+\triangleq\{i: p_{1i}-p_{2i}> 0\}$ and $\mathcal{I}_-\triangleq\{i: p_{1i}-p_{2i}\leq 0\}$. The following theorem give optimal solutions of the EIT approximation in \eqref{eq:ML_HPapproximation} for $l\leq 1$. 

\begin{theorem}\label{Theorem:ML_HP_approximation}
   For non-binary sources in the high privacy regime, i.e., $l\in [0,1]$, the maximal utility of the EIT approximation \eqref{eq:ML_HPapproximation} is
	\begin{align}
		\frac{(2^l-1)}{2}\|\bp_1-\bp_2\|^2_1.
	\end{align} 
	The optimal mechanism $\bW^*$ has two unique columns: one of the two columns has nonzero entries given by $2^l-1$ only in the positions indexed by $\mathcal{I}_+$ while the other column has the same nonzero entries only in the positions indexed by $\mathcal{I}_-$. Each of the remaining $M-2$ columns of $\bW^*$ has the same entry in each row and the sum of these $M-2$ entries is $2-2^l$.
\end{theorem}
The proof of Theorem \ref{Theorem:ML_HP_approximation} is in Appendix \ref{proof:Theorem:ML_HP_approximation}. The proof hinges on the following two simplifications:
\begin{itemize}
	\item[(i)] Since $\bp_1-\bp_2$ contains both positive and negative entries, maximizing $\|(\bp_1-\bp_2)\bW\|$ requires that every column of the optimal solution $\bW^*$ has either the maximal and minimal values of the column in the positions indexed by $\mathcal{I}_+$ and $\mathcal{I}_-$, respectively, or vice versa. This allows finding the structure of $\bW^*$.
	\item[(ii)] We further exploit  the EIT approximation   that all rows of the optimal mechanism $\bW^*$ are in a small neighborhood centered at $\bw_0$ to find the optimal $\bw_0$. This relies on (i) above in exploiting the structure of $\bW^*$.
\end{itemize}
Thus, Theorem \ref{Theorem:ML_HP_approximation} shows that to preserve the utility of binary hypothesis testing while matching the maximal leakage, the privacy mechanism first splits all input symbols into two subsets $\mathcal{S}_1$ and $\mathcal{S}_2=\mathcal{X}\setminus\mathcal{S}_1$. Specifically, every symbol in $\mathcal{S}_1$ has a higher probability under $H_1$ than under $H_2$, i.e., the indices of symbols in $\mathcal{S}_1$ (resp. $\mathcal{S}_2$) are in $\mathcal{J}_+$ (resp. $\mathcal{J}_-$). The privacy mechanism then maps all symbols of $\mathcal{S}_1$ with the same probability to a single output. Similarly, all symbols in $\mathcal{S}_2$ are mapped with the same probability to a single output that is distinct from that of $\mathcal{S}_1$. Therefore, observing one of these two output symbols, we know the corresponding input subset even if we cannot identify the exact input symbols within the subset.
\begin{remark}
	From Theorem \ref{Theorem:ML_HP_approximation}, we have that if either $\mathcal{I}_+$ or $\mathcal{I}_-$ has only one element, the privacy mechanism $\bW^*$ of the EIT approximation will reveal one input symbol as is. 
\end{remark}

\subsection{Linear Approximation in High Utility Regime}
From Lemma \eqref{Lemma:Maximal_Leakage}, recall that for the perfect utility, i.e., $l=\log M$, all column permutations of the identity matrix are optimal. Without loss of generality, we choose the identity matrix, i.e., $\bW=\mathbf{I}$, as the perfect utility achieving mechanism. In the high utility regime, i.e., $l\geq \log(M-1)$, by expanding the Taylor series around the identity matrix, we can approximate the objective to the first order as $D(\bp_1\bW\|\bp_2\bW)= \text{Tr}(\boldsymbol{\Psi}\bW^T)+o(\|(\bp_1-\bp_2)\bW\|^2)$, where the expression $\text{Tr}(\boldsymbol{\Psi}\bW^T)$ represents the trace of the matrix $\boldsymbol{\Psi}\bW^T$ and $\bPsi$ is the partial derivative matrix of $D(\bp_1\bW\|\bp_2\bW)$ calculated at the identity matrix, and has entries equal to
\begin{align}\label{eq:partial_KLDivergence}
\Psi_{ij}
=p_{1i}\Big(\log\frac{p_{1j}}{p_{2j}}+\log e\Big)-p_{2i}\Big(\frac{p_{1j}}{p_{2j}}\log e\Big).
\end{align}
The resulting PUT in the high utility regime is	
\begin{equation}\label{eq:ML_HUapproximation}
\begin{aligned}
\max_{\substack{\bW}} \quad & \text{Tr}(\boldsymbol{\Psi}\bW^T)\\
\text{s.t.} \quad 
& \sum_{j=1}^{M} \max_{i} W_{ij}\leq 2^l\\
&\sum_{j=1}^{M}W_{ij}=1 \qquad\, \text{for all }i\\
&  W_{ij}\geq 0 \qquad\qquad \text{for all }i,j\\
\end{aligned}
\end{equation}
where $l\geq \log(M-1)$. The optimization in \eqref{eq:ML_HUapproximation} is a linear program and can be efficiently solved. The following lemma presents a property of the optimal solutions of \eqref{eq:ML_HUapproximation}.
\begin{lemma}\label{Lemma:Solu_LinearApprox}
For the high utility regime, i.e., $l\geq \log (M-1)$, the optimal solution $\bW^*$ of \eqref{eq:ML_HUapproximation} has no less than $M(M-2)$ zero entries such that all diagonal entries are positive and every input symbol is mapped to at most two output symbols.
\end{lemma}
The proof of Lemma \ref{Lemma:Solu_LinearApprox} is in Appendix \ref{proof:Lemma:Solu_LinearApprox}. 
Basically, the proof follows from the fact that the maximal entry in each row of $\bPsi$ in \eqref{eq:partial_KLDivergence} is on the diagonal and that the objective in \eqref{eq:ML_HUapproximation} is the sum of convex combinations of every row of $\bPsi$. Thus, in the high utility regime, at least one revealed symbol reduces uncertainty about the input to at most one bit!

\section{Concluding Remarks}
We have developed the PUTs for the hypothesis testing problem using ML as the privacy metric and the error exponent as the utility metric. Our results for both binary and $M$-ary data suggest that the mechanism guaranteeing a bounded ML, i.e., limiting guesses about arbitrary functions of $X^n$, is able to maximize the utility only by {\em partially} or {\em fully revealing} a few input symbols (even in the high privacy regime). 
This raises concern about the appropriateness of ML for this problem. 
In contrast, in the high privacy regime, the mutual information privacy metric~\cite{Liao_Allerton16} yields output distributions with {\em little or no semblance} to the original distributions for \emph{all} input symbols. 

\appendix
\subsection{Proof of Lemma \ref{Lemma:Maximal_Leakage}}\label{proof:Lemma:Maximal_Leakage}
\begin{proof}
	From the row stochasticity condition of the conditional probability matrix $\bW$, we have
	\begin{align}
	&0 \leq W_{ij}\leq 1\\
	&\sum_{j=1}^{M}W_{ij}=1 \text{ for all } j,
	\end{align}
	directly from which, the upper and lower bounds of the sum of the maximal value of every column of $\bW$ are
	\begin{align}
	1 \leq \sum_{j=1}^{M}\max_i\{W_{ij}\} \leq M\cdot 1=M,
	\end{align}
	and then, the function $I_{\infty}(\bp,\bW)$ is bounded by
	\begin{align}
	0\leq I_{\infty}(\bp,\bW)\leq \log M,
	\end{align}
	where the left equality holds if and only if every column of $\bW$ has the same value, i.e., for each $j\in\{1,...,M\}$, $\max_i\{W_{ij}\}=W_{ij}$ for all $i$, thus, the $\bW$ is a rank-1 matrix; and the right equality holds if and only if the maximal value of every column is 1, i.e., $\max_i\{W_{ij}\}=1$ for all $j$, and therefore, $\bW$ is a permutation of identity matrix $\mathbf{I}$.
\end{proof}

\subsection{Proof of Theorem \ref{theorem:ML_OptSolChara}}\label{proof:theorem:ML_OptSolChara}
\begin{proof}
	From the constraints in \eqref{eq:MaximalLeakage_original}, any column permutation of a feasible $\bW$ is also feasible since the unchanged leakage and row stochasticity conditions are satisfied. Let $\bW^*$ be an optimal solution of \eqref{eq:MaximalLeakage_original} achieving
	\begin{align}
	D(\bp_1\bW^*\|\bp_2\bW^*)=\sum_{j=1}^{M}\bp_1\bW^*_j
	\log\frac{\bp_1\bW^*_j}{\bp_2\bW^*_j}
	\end{align}
	where $\bW_j^*$ is the $j^{\text{th}}$ column of $\bW^*$. Since addition is commutative, thus any column permutation of $\bW^*$ gives the same objective value, i.e., any column permutation of $\bW^*$ is also an optimal solution of \eqref{eq:MaximalLeakage_original}.
	
	If $\bW^*$ has one zero column, we can generate a different optimal solution $\bar{\bW}^*$ by permuting the zero column with another non-zero column. Without loss of generality, assume all entries of the first column of $\bW^*$ are zero and $\bar{\bW}^*$ is generated by permuting the first two columns of $\bW^*$. Let $\bW^*_j$ and $\bar{\bW}^*_j$, for all $j\in\{1,...,M\}$, denote the $j^{\text{th}}$ columns of $\bW^*$ and $\bar{\bW}^*$, respectively. Thus, $\bW^*_1=\bar{\bW}^*_2=\mathbf{0}$ and $\bW^*_2=\bar{\bW}^*_1$.
	The feasible region of \eqref{eq:MaximalLeakage_original} is convex, thus all convex combinations of $\bW^*$ and $\bar{\bW}^*$, indicated as $\lambda\bW^*+(1-\lambda)\bar{\bW}^*$, $\lambda\in[0,1]$, are also feasible. The objective value of $\lambda\bW^*+(1-\lambda)\bar{\bW}^*$ can be written as
	\begin{align}
	&D(\bp_1(\lambda\bW^*+(1-\lambda)\bar{\bW}^*)\|\bp_2(\lambda\bW^*+(1-\lambda)\bar{\bW}^*))\nonumber\\
	=&\sum_{j=1}^{M}\bp_1\big(\lambda\bW^*_j+(1-\lambda)\bar{\bW}^*_j\big)\log \frac{\bp_1\big(\lambda\bW^*_j+(1-\lambda)\bar{\bW}^*_j\big)}{\bp_2\big(\lambda\bW^*_j+(1-\lambda)\bar{\bW}^*_j\big)}\nonumber\\
	=&(1-\lambda)\bp_1\bar{\bW}^*_1\log \frac{(1-\lambda)\bp_1\bar{\bW}^*_1}{(1-\lambda)\bp_2\bar{\bW}^*_1}+\lambda\bp_1\bW^*_2\log \frac{\lambda\bp_1\bW^*_2}{\lambda\bp_2\bW^*_2}\nonumber\\
	&+\sum_{j=3}^{M}\bp_1\bW^*_j\log \frac{\bp_1\bW^*_j}{\bp_2\bW^*_j}\\
	=&\sum_{j=2}^{M}\bp_1\bW^*_j\log \frac{\bp_1\bW^*_j}{\bp_2\bW^*_j}=D(\bp_1\bW^*\|\bp_2\bW^*)\nonumber. 
	\end{align}
	That is, all convex combinations of $\bW^*$ and $\bar{\bW}^*$ are optimal. Thus, an infinite number of solutions 
	are optimal.
\end{proof}

\subsection{Proof of Theorem \ref{theorem:ML_binarysource}}\label{proof:theorem:ML_binarysource}
\begin{proof}
	For binary sources, \textit{privacy mechanism} $\bW$ is a $2\times 2$ matrix, which can be expressed as 
	\begin{align}
	\bW=\begin{bmatrix}
	1-\rho_1 & \rho_1\\
	\rho_2 &1-\rho_2
	\end{bmatrix}
	\end{align}
	where $1 \geq \rho_1,\rho_2\geq 0$. Then,
	\begin{equation}
	\sum_{j=1}^{2}\max_iW_{ij}=
	\begin{cases}
	2-\rho_1-\rho_2, \quad \rho_1+\rho_2<1\\
	\rho_1+\rho_2,   \quad\qquad \rho_1+\rho_2>1
	\end{cases}
	\end{equation}
	Therefore, the privacy constraint in \eqref{eq:MaximalLeakage_Constraint} simplifies as
	\begin{align}
	2-2^l\leq \rho_1+\rho_2\leq 2^l, 
	\end{align}
	and along with the non-negativity of entries of $\bW$, requiring $\rho_1,\rho_2\geq 0$, yield the feasible region shown as the shaded region in Fig.~\ref{fig:ML_binaryregion}.
	
	The original PUT \eqref{eq:MaximalLeakage_original} maximizes the convex function $D(\bp_1\bW\|\bp_2\bW)$ over the shaded polytope as shown in Fig. \ref{fig:ML_binaryregion}, thus at least one of the corner points of the polytope is optimal, i.e., an optimal solution is one of the vertexes \raisebox{.5pt}{\textcircled{\raisebox{-.9pt} {$1$}}\--\textcircled{\raisebox{-.9pt} {$6$}}} in Fig. \ref{fig:ML_binaryregion}. The utilities at vertexes \raisebox{.5pt}{\textcircled{\raisebox{-.9pt} {$5$}} and \textcircled{\raisebox{-.9pt} {$6$}}} are 0. The vertexes \raisebox{.5pt}{\textcircled{\raisebox{-.9pt} {$1$}} \big(resp. \textcircled{\raisebox{-.9pt} {$2$}}\big)} and \raisebox{.5pt}{\textcircled{\raisebox{-.9pt} {$4$}} \big(resp. \textcircled{\raisebox{-.9pt} {$3$}}\big)} are from column permutations, and thus, referring to Theorem \ref{theorem:ML_OptSolChara}, \raisebox{.5pt}{\textcircled{\raisebox{-.9pt} {$1$}} \big(resp. \textcircled{\raisebox{-.9pt} {$2$}}\big)} and \raisebox{.5pt}{\textcircled{\raisebox{-.9pt} {$4$}} \big(resp. \textcircled{\raisebox{-.9pt} {$3$}}\big)} have the same utility. Thus, to get the optimal value, it suffices to compare the utilities of \raisebox{.5pt}{\textcircled{\raisebox{-.9pt} {$1$}} and \textcircled{\raisebox{-.9pt} {$2$}}}.
	\begin{itemize}
		\item[1.] The \textit{privacy mechanism} $\bW$ at \raisebox{.5pt}{\textcircled{\raisebox{-.9pt} {$1$}}} is 
		\begin{align}
		\begin{bmatrix}
		2-2^l & 2^l-1\\
		1 & 0
		\end{bmatrix},
		\end{align} and then, the utility of \raisebox{.5pt}{\textcircled{\raisebox{-.9pt} {$1$}}} is given by \eqref{eq:binary_optimalvalue_point1}.
		\item[2.] The \textit{privacy mechanism} $\bW$ at \raisebox{.5pt}{\textcircled{\raisebox{-.9pt} {$2$}}} is 
		\begin{align}
		\begin{bmatrix}
		0 & 1\\
		2^l-1 & 2-2^l
		\end{bmatrix},
		\end{align} and then, the utility of \raisebox{.5pt}{\textcircled{\raisebox{-.9pt} {$2$}}} is presented in \eqref{eq:binary_optimalvalue_point2}.
	\end{itemize}
	Therefore, for $M=2$ the optimal utility of \eqref{eq:MaximalLeakage_original} is the maximal value of \eqref{eq:binary_optimalvalue_point1} and \eqref{eq:binary_optimalvalue_point2}.	
\end{proof}

\subsection{Proof for Theorem \ref{Theorem:ML_HP_approximation}}\label{proof:Theorem:ML_HP_approximation}
\begin{proof}
	The objective in \eqref{eq:ML_HPapproximation} can be bounded as
	\begin{subequations}
		\begin{align}
	&\big\|(\mathbf{p}_1-\mathbf{p}_2)\bW[(\mathbf{w}_0)^{-\frac{1}{2}}]\big\|^2=\sum_{j=1}^{M}\frac{\big((\mathbf{p}_1-\mathbf{p}_2)\bW_j\big)^2}{w_{0j}}\nonumber\\
			=&\sum_{j=1}^{M}\frac{\big(\sum_{i\in\mathcal{I}_+}(p_{1i}-p_{2i})W_{ij}+\sum_{i\in\mathcal{I}_-}(p_{1i}-p_{2i})W_{ij}\big)^2}{w_{0j}}\\
			\label{eq:ML_HPTheoremproof_inequality}
			\leq & \sum_{j=1}^{M}\bigg(\frac{\max_iW_{ij}\sum_{i\in\mathcal{I}_+}(p_{1i}-p_{2i})}{\sqrt{w_{0j}}}\\
			&\qquad\quad +\frac{\min_iW_{ij}\sum_{i\in\mathcal{I}_-}(p_{1i}-p_{2i})}{\sqrt{w_{0j}}}\bigg)^2\nonumber\\	
			\label{eq:ML_HPTheoremproof_equality}
			=&\frac{1}{4}\|\bp_1-\bp_2\|^2_1\sum_{j=1}^{M}\frac{(\max_iW_{ij}-\min_iW_{ij})^2}{w_{0j}}
		\end{align}
	\end{subequations}
	where the inequality \eqref{eq:ML_HPTheoremproof_equality} directly results from
	\begin{align}
		\bigg|\sum_{i\in\mathcal{I}_-}(p_{1i}-p_{2i})\bigg|=\bigg|\sum_{i\in\mathcal{I}_+}(p_{1i}-p_{2i})\bigg|=\frac{1}{2}\|\bp_1-\bp_2\|_1.\nonumber
	\end{align}
	Let $W_{j,\max}$ and $W_{j,\min}$ be the maximal and minimal values of the $j^{\text{th}}$ column of $\bW$, respectively.
	The sufficient and necessary condition for the equality of \eqref{eq:ML_HPTheoremproof_inequality} is that the $j^\text{th}$ column of $\bW$, for all $j\in\{1,\ldots,M\}$, has entries as either	
	\begin{align}
		\label{eq:ML_HPTheoremproof_W1}
		W_{ij}=&\begin{cases}
			W_{j,\max}\quad \text{for } i\in\mathcal{I}_+ \\
			W_{j,\min}\quad \text{for } i\in\mathcal{I}_- \\
		\end{cases}
	\end{align}
	or
	\begin{align}
		\label{eq:ML_HPTheoremproof_W2}
		W_{ij}=&\begin{cases}
			W_{j,\min}\quad \text{for } i\in\mathcal{I}_+ \\
			W_{j,\max}\quad \text{for } i\in\mathcal{I}_- \\
		\end{cases}
	\end{align}
	We now show that the optimal solution $\bW^*$ of \eqref{eq:ML_HPapproximation} has at most two columns for which $W^*_{j,\max}-W^*_{j,\min}>0$. For $l>0$, i.e., $2^l>1$, since $\bW^*$ achieves the maximal leakage of $l$ bits and is row stochastic, the maximal values of the columns of $\bW^*$ are not all in the same row. Thus, $\bW^*$ has at least two columns, one in the form of \eqref{eq:ML_HPTheoremproof_W1} and the other in the form of \eqref{eq:ML_HPTheoremproof_W2}. The remaining columns of $\bW^*$ are either in the form of \eqref{eq:ML_HPTheoremproof_W1} or of \eqref{eq:ML_HPTheoremproof_W2}. Let $\mathcal{J}_1=\big\{j: W^*_{ij} \text{ is in the form of } \eqref{eq:ML_HPTheoremproof_W1}\}$ and $\mathcal{J}_2=\big\{j: W^*_{ij} \text{ is in the form of } \eqref{eq:ML_HPTheoremproof_W2}\big\}$. Thus, we have
	\begin{subequations}
		\begin{align}
			\label{eq:ML_HP_inproof1}
			&\sum_{j=1}^{M}W^*_{j,\max}= 2^l\\
			\label{eq:ML_HP_inproof2}
			&\sum_{j\in\mathcal{J}_1}W^*_{j,\max}+\sum_{j\in\mathcal{J}_2}W^*_{j,\min}=1\\
			\label{eq:ML_HP_inproof3}
			&\sum_{j\in\mathcal{J}_1}W^*_{j,\min}+\sum_{j\in\mathcal{J}_2}W^*_{j,\max}=1.
		\end{align}
	\end{subequations}
	From \eqref{eq:ML_HP_inproof1}-\eqref{eq:ML_HP_inproof3}, we get
	\begin{align}
		&\sum_{j=1}^{M}(W^*_{j,\max}-W^*_{j,\min})=2^{l+1}-2	
	\end{align}
	Furthermore, subtracting \eqref{eq:ML_HP_inproof2} or \eqref{eq:ML_HP_inproof3} from \eqref{eq:ML_HP_inproof1} and observing that $W^*_{j,\max}-W^*_{j,\min}\geq 0$ for all $j$, we have
	\begin{align}
		&\sum_{j\in\mathcal{J}_k}(W^*_{j,\max}-W^*_{j,\min})=2^l-1\quad \text{ for } k=1,2\\
		\label{eq:ML_HP_inproof4}
		\Rightarrow \;\; & W^*_{j,\max}-W^*_{j,\min}\leq 2^l-1 \quad j\in\mathcal{J}_k,\,k=1,2,
	\end{align}
	where the equality in \eqref{eq:ML_HP_inproof4} holds if and only if $\bW^*$ has at most two columns, indexed by $j_1^*\in \mathcal{J}_1$ and $j_2^*\in \mathcal{J}_2$ such that $\bW^*_{j_k^*,\max}-\bW^*_{j_k^*,\min}>0$, $k=1,2$, which also means that for the remaining $M-2$ columns of $\bW^*$, $\bW^*_{j,\max}-\bW^*_{j,\min}=0$, $j\in\{1,\ldots,M\}\setminus\{j_1^*,j_2^*\}$.
	Without loss of generality, let $j^*_1=1$, $j^*_2=2$, $\mathcal{I}_+=\{i \in\{1,\ldots,m\}, m\leq M-1\}$ and $\mathcal{I}_-=\{i \in\{m+1,\ldots,M\}\}$. Thus, the first $m$ entries of the first column of $\bW^*$ are $W_{1,\min}+2^l-1$ while the remaining entries are $W_{1,\min}$. Similarly, for the second column, the first $m$ entries are $W_{2,\min}$ while the remaining entries are $W_{2,\min}+2^l-1$. Finally, the $i^{\text{th}}$ column has the same entry $\epsilon_{i-2}\geq 0$ for $i\in\{3,\ldots,M\}$. One can verify that this simplification leads to two unique rows in $\bW^*$ given by $[W_{1,\min}+2^l-1,W_{2,\min},\epsilon_1,\epsilon_2,\ldots,\epsilon_{M-2}]$ and $[W_{1,\min},W_{2,\min}+2^l-1,\epsilon_1,\epsilon_2,\ldots,\epsilon_{M-2}]$. All column permutations of $\bW^*$, i.e., permuting the two rows above simultaneously, are permitted.  
	
	From \eqref{eq:ML_HPTheoremproof_equality}, we deduce that $w_{0j}$ is relevant only for $j=1,2$ and we now use the EIT approximation to determine these values.
    Since the EIT approximation requires that all rows of the $\bW^*$ are in a ball of radius $\delta$ about $\bw_0$, we can find the corresponding $\bw_0$ as the average of the two unique rows of $\bW^*$ to satisfy the neighborhood condition. Due to the fact that entries of the first two columns are either $W^*_{k,\min}$ or $W^*_{k,\min}+2^l-1$, $k=1,2$, the first two entries of the $\bw_0$ are $\frac{2W^*_{1,\min}+2^l-1}{2}$ and $\frac{2W^*_{2,\min}+2^l-1}{2}$. Therefore, the corresponding optimal value of~\eqref{eq:ML_HPapproximation} is
	\begin{align}
	\nonumber
		\frac{(2^l-1)^2\|\bp_1-\bp_2\|^2_1}{4}\sum_{k=1}^{2}\frac{1}{2W^*_{k,\min}+2^l-1},
	\end{align}
	which in turn is maximized by $W^*_{k,\min}=0$, $k=1,2$. Thus, the optimal privacy mechanism is
	\begin{align}
	\bW^*=&\begin{bmatrix}
	2^{l}-1 & 0 & \epsilon_1 & \epsilon_2 &\dots &\epsilon_{M-2}\\
	\vdots & \vdots & \vdots & \vdots & \ddots  &\vdots\\
	2^{l}-1 & 0 & \epsilon_1 & \epsilon_2 &\dots &\epsilon_{M-2}\\
	0 & 2^{l}-1 & \epsilon_1 & \epsilon_2 &\dots &\epsilon_{M-2}\\
	\vdots   & \vdots  & \vdots & \vdots &\ddots  & \vdots \\
	0 & 2^{l}-1 & \epsilon_1 & \epsilon_2 &\dots &\epsilon_{M-2}\\
	\end{bmatrix}
	\end{align}
	where the first $m$ rows are the same and the remaining $M-2$ rows are the same. Note that the non-negative $\epsilon_i$, $i\in\{1,\ldots,M-2\}$, sum up to $2-2^l$.
\end{proof}

\subsection{Proof of Lemma \ref{Lemma:Solu_LinearApprox}}\label{proof:Lemma:Solu_LinearApprox}
\begin{proof}
	From \eqref{eq:partial_KLDivergence}, we know that diagonal entries of the partial derivative matrix $\bPsi$ is
	\begin{align}
	\Psi_{ii}=p_{1i}\log\frac{p_{1i}}{p_{2i}}\quad \text{for all } i\in\{1,...,M\}.
	\end{align}
	In addition, for all $i,j\in\{1,...,M\}$, we have
	\begin{align}
	&\Psi_{ii}-\Psi_{ij}\nonumber\\
	=&p_{1i}\log\frac{p_{1i}}{p_{2i}}-p_{1i}\Big(\log\frac{p_{1j}}{p_{2j}}+\log e\Big)+p_{2i}\Big(\frac{p_{1j}}{p_{2j}}\log e\Big)\nonumber\\
	=&-p_{1i}\log\frac{p_{2i}p_{1j}}{p_{1i}p_{2j}}+p_{1i}\log e\Big(-1+\frac{p_{2i}p_{1j}}{p_{1i}p_{2j}}\Big)\nonumber\\
	\geq &-p_{1i}\Big(\frac{p_{2i}p_{1j}}{p_{1i}p_{2j}}-1\Big)\log e +p_{1i}\Big(\frac{p_{2i}p_{1j}}{p_{1i}p_{2j}}-1\Big)\log e =0.
	\end{align}
	That is to say, for every row of the partial derivative matrix $\bPsi$, the maximal value is on the diagonal.
	
	Let $\bW^*$ be the optimal solution of \eqref{eq:ML_HUapproximation} and positive integers $n_i$, $i\in\{1,...,M\}$, indicate the number of nonzero entries of the $i^{\text{th}}$ row in $\bW^*$. Since all rows of $\bW^*$ sum up to 1, the objective in \eqref{eq:ML_HUapproximation} is the sum of convex combinations of every row of $\bPsi$. Therefore, while satisfying the ML constraint, maximizing $\text{Tr}(\boldsymbol{\Psi}\bW^T)$ requires that choosing the diagonal entries of $\bW^*$ as large as possible, and the convex combination of the $i^\text{th}$ row of $\bPsi$ involves the $n_i$ maximal entries of the row. In general, for all $i\in\{1,...,M\}$, $n_i$ should be as small as possible. Specially, for $l=\log M$, the optimal value of \eqref{eq:ML_HUapproximation} is $D(\bp_1\|\bp_2)$ such that all diagonal entries of $\bW^*$ should be $1$ and $n_i=1$ for all $i$, i.e., $\bW^*$ is the identity matrix.
	
	For $l\geq \log (M-1)$, we have $\sum_{j=1}^{M}\max_i W^*_{ij}=2^l\geq M-1$. Construct a matrix $\bW$ such that every row of $\bW$ has only two non-zero entries at positions indexed by the first and second maximal entries of the corresponding row of $\bPsi$. Specifically, all diagonal entries of $\bW$ are $\frac{2^l}{M}$ ($\frac{2^l}{M}\geq \frac{M-1}{M}$) and the another non-zero entry of each row in $\bW$ is $\frac{M-2^l}{M}$ ($\frac{M-2^l}{M}\leq \frac{1}{M}$). The $\bW$ is feasible and $n_i$ can be at least as small as $2$ for all $i$. Therefore, since the number of zero entries of $\bW$ is $(M-2)M$, the number of zero entries of $\bW^*$ is no less than $(M-2)M$.
\end{proof}

\bibliographystyle{IEEEtran}
\bibliography{HypothesisTestingML}
\end{document}